\documentstyle[12pt]{article}
\def\pw#1{^{#1}}

\def\beeq{\begin{eqnarray}} \def\eeeq{\end{eqnarray}}
\newcommand\mysection{\setcounter{equation}{0}\section}
\renewcommand{\theequation}{\thesection.\arabic{equation}}
\newcounter{hran} \renewcommand{\thehran}{\thesection.\arabic{hran}}

\def\bmini{\setcounter{hran}{\value{equation}}
  \refstepcounter{hran}\setcounter{equation}{0}
  \renewcommand{\theequation}{\thehran\alph{equation}}\begin{eqnarray}}

\def\bminiG#1{\setcounter{hran}{\value{equation}}
\refstepcounter{hran}\setcounter{equation}{-1}
\renewcommand{\theequation}{\thehran\alph{equation}}
\refstepcounter{equation}\label{#1}\begin{eqnarray}}

\def\emini{\end{eqnarray}\relax\setcounter{equation}{\value{hran}}\renewcommand{\theequation}{\thesection.\arabic{equation}}}

\def\cO#1{{\cal{O}}\!\left(#1\right)}

\def\fun#1#2{\lower3.6pt\vbox{\baselineskip0pt\lineskip.9pt
  \ialign{$\mathsurround=0pt#1\hfil##\hfil$\crcr#2\crcr\sim\crcr}}}

\def\eV{{\rm e\kern-0.12em V}}            
  
\def\half{{\textstyle {1\over2}}}
  \def\quart{{\textstyle {1\over4}}}

\def \l {{\ell_{\kappa}}}

\def \al {\relax\ifmmode{\alpha}\else{$\alpha${ }}\fi}
    \def\Re{\mathop{\rm Re}}
\def\abs#1{\left| #1\right|}
\def\lrang#1{\left\langle #1 \right\rangle}

\def\ben{\begin{enumerate}}  \def\een{\end{enumerate}}
\def\bit{\begin{itemize}}    \def\eit{\end{itemize}}
\def\beq{\begin{equation}}   \def\eeq{\end{equation}}
\def\bea{\begin{eqnarray}}  \def\eea{\end{eqnarray}}
\def\nn{\nonumber}

\def\kp{\relax\ifmmode{k_\perp}\else{$k_\perp${ }}\fi}
\def\kps{\relax\ifmmode{k_\perp\pw2}\else{$k_\perp\pw2${ }}\fi}
\def \as{\relax\ifmmode\alpha_s\else{$\alpha_s${ }}\fi}

\def\br{brems\-strah\-lung\ }    

\newskip\humongous \humongous=0pt plus 1000pt minus 1000pt
\def\caja{\mathsurround=0pt}
\def\eqalign#1{\,\vcenter{\openup1\jot
\caja   \ialign{\strut \hfil$\displaystyle{##}$&$
\displaystyle{{}##}$\hfil\crcr#1\crcr}}\,}

\newif\ifdtup

\def\eqal2#1{\,\vcenter{\openup1\jot
\caja   \ialign{\strut \hfil$\displaystyle{##}$&\hfil$
\displaystyle{{}##}$\hfil &$
\displaystyle{{}##}$\hfil\crcr#1\crcr}}\,}

 \def\cite#1{[\ref{#1}]}
 \def\citd#1#2{[\ref{#1},\ref{#2}]}
 \def\citt#1#2#3{[\ref{#1},\ref{#2},\ref{#3}]}

\def\vq#1{\vec{q}_{#1}}

\def\beql#1{\beq\label{#1}}
\def\beal#1{\bea\label{#1}}

\def\np#1#2#3{{\em Nucl.~Phys.}~\underline{B#1} (19#3) #2}
\def\pl#1#2#3{{\em Phys.~Lett.}~\underline{#1B} (19#3) #2}
\def\prD#1#2#3{{\em Phys.~Rev.}~\underline{D#1} (19#3) #2}
\def\pr#1#2#3{{\em Phys.~Rev.}~\underline{#1} (19#3) #2}

\def\prl#1#2#3{{\em Phys.Rev.Lett.}~\underline{#1} (19#3) #2}

\def\sjnp#1#2#3{{\em Sov.J.Nucl.Phys.}~\underline{#1} (19#3) #2}
\def\spj#1#2#3{{\em Sov.Phys.JETP}\/~\underline{#1} (19#3) #2}
\def\spd#1#2#3{{\em Sov.Phys.Dokl.}\/~\underline{#1} (19#3) #2} 
\def\spu#1#2#3{{\em Sov.Phys.Usp.}\/~\underline{#1} (19#3) #2}

\def\zp#1#2#3{{\em Z.~Phys.}~\underline{C#1} (19#3) #2}

\begin{document}

\thispagestyle{plain}
\setcounter{page}{1}
\vbox to 1 truecm {}

\begin{flushright}
February 1996 \\[0.1cm]

BI-TP 95-40\\
CERN-TH.96/14\\
CUTP-724\\
LPTHE-Orsay 95-84
 \\
\end{flushright}

\vfill
\def\cen{\centerline}

\cen{{\bf\large THE LANDAU-POMERANCHUK-MIGDAL}}
\cen{{\bf\large EFFECT IN QED }}
\renewcommand{\thefootnote}{\fnsymbol{footnote}}
\vskip 1.5 truecm
\centerline
{\bf R.~Baier~$\pw1$, Yu.~L.~Dokshitzer~$\pw2$, A.~H.~Mueller\footnote[1]{Supported in  part by the U.S. Department of Energy under grant DE-FG02 94ER 40819}~$\pw3$, 
S.~Peign\'e~$\pw4$ and D.~Schiff~$\pw4$}
\vskip 10 pt
\centerline{{\it $\pw{1 }$Fakult\"at f\"ur Physik,
Universit\"at Bielefeld, D-33501 Bielefeld, Germany}}
\centerline{{\it $\pw{2 }$ Theory Division, CERN, 1211 Geneva 23, 
Switzerland\footnote[2]
{Permanent address: Petersburg Nuclear Physics Institute, Gatchina, 
 188350 St. Petersburg, Russia } }}
\centerline{{\it $\pw{3 }$Physics Department, Columbia University, 
New York, NY 10027, USA}}
\centerline{{\it $\pw{4 }$
LPTHE\footnote[3] { Laboratoire associ\'e au
Centre National de la Recherche Scientifique}
, Universit\'e  Paris-Sud, B\^atiment 211, F-91405 Orsay, France}}
 
\renewcommand{\thefootnote}{\arabic{footnote}}
\vskip 2 cm
\cen{\bf Abstract}
\vskip 5pt
\noindent
{The calculation of the radiative energy loss encountered by a fast charged 
particle which undergoes multiple scattering is being investigated. 
A detailed derivation of the Landau-Pomeranchuk-Migdal coherent effect 
in QED is given, focusing on the specific feature of the Coulomb interaction. 
As a result the radiation intensity per unit length in the coherent regime 
is shown to be proportional to $\sqrt \omega$ for a photon energy $\omega$ 
times a logarithmic enhancement which is determined exactly.}

\vfill \eject

\mysection{Introduction}
\label{sec:nom1}
The radiative energy loss encountered by a charged fast particle which
undergoes successive scatterings in a medium has recently been the object of
renewed interest. This is indeed of special importance in order to understand
the energy loss mechanism for quarks and gluons, propagating in nuclear matter
and in particular through the quark-gluon plasma. \par

Our first step \cite{BDPS} has been to reformulate the QED treatment given 
by Landau, Pome\-ranchuk and Migdal (LPM) \citt{LP}{M1}{Ter}. 
The experimental observation of the LPM effect at SLAC \cite{SLAC} has
recently triggered further studies with special emphasis on finite length 
effects \cite{BD}. \par

In the present paper we focus on a detailed derivation of the radiative 
spectrum per unit length, assuming a very large number of
scatterers. The model used which has been recently elaborated by Gyulassy 
and Wang \cite{GW}
depicts the multiple scattering of a fast electron in the medium as due to
static scattering centres with Debye screened Coulomb potentials. The
assumption that the mean free path $\lambda$ of the projectile is much larger
than the screening radius $\lambda \gg \mu^{-1}$ allows one to treat successive
scatterings as independent. Coupled to the soft photon approximation this 
then leads
to an eikonal picture of classical propagation. \par

As stated in \cite{BDPS}, 
the specific case of the Coulomb potential 
which is not screened at {\em short}\/ distances requires 
a special treatment which was not available in the literature. 
Such a treatment is essential for QCD. 
We present here for QED a complete detailed derivation, correcting 
an unjustified step made in \cite{BDPS}. \par
For the specific case of the Coulomb interaction, the radiation intensity 
per unit length in the coherent regime is proportional to $\sqrt \omega$ 
times a logarithmic enhancement. For potentials less singular than 
the Coulomb one at short distance, where the random walk picture 
is applicable, we confirm the original LPM result. 

The outline of the paper is as follows~: \\
In section \ref{sec:nom2}, we show in details how the model of static 
Coulomb centres describes multiple scattering. 
The radiation spectrum induced by multiple scattering
is worked out in the soft photon approximation in section \ref{sec:nom3} 
and studied in section \ref{sec:nom4}. 
In this last section, we first obtain the shape of the 
spectrum on heuristic grounds and then carefully derive the spectrum in the 
coherent regime by solving an exact differential equation to leading 
logarithmic accuracy. Section \ref{conc} is the conclusion.

\mysection{Model for multiple scattering}
\label{sec:nom2}
In order to work out the radiation intensity properly normalized we need first
to derive the multiple scattering cross section of a fast electron propagating
in a medium. We do this using the model of Gyulassy and Wang \cite{GW}. 
The main feature of this model consists in assuming that scattering centres 
are static. Thus the collisional energy loss of the charged particle 
vanishes, and the total energy loss will be purely radiative. 
The scattering centre located at 
$\vec{x}_i$ 
creates screened Coulomb potential
\beql{potential1}
{\cal{V}}_i(\vec{x}) = {e \over 4 \pi} {e^{-\mu|\vec{x} - \vec{x}_i|}
\over |\vec{x} - \vec{x}_i|}\>,
\eeq
with Fourier transform
\beql{potential2}
{\cal{V}}_i(\vec{q}) 
= {e \over \vec{q}\ ^2 + \mu^2} e^{-i\vec{q}\cdot \vec{x}_i}\>.
\eeq
Here $\mu$ is the Debye mass in the medium and we denote 
$\vec{q}_i$ the momentum transferred to the electron
during scattering on 
the 
centre $\vec{x}_i$.

To be able to treat successive elastic scatterings as independent, centres 
have to be well separated. This means that the average distance between two 
successive interactions (the electron mean free path $\lambda$), is large 
compared to the range of the potential,
\beql{ind.scatt.}
\lambda \gg \mu^{-1}\>.
\eeq
This leads to important simplifications in the study of the radiative energy 
loss.

We also concentrate on the high energy limit, where the transferred momenta 
$\vq{i}$ become transverse. 
Moreover, their characteristic values are of order  $\mu$, and we have
\beql{smallangle}
\Theta_s ={|\vec{q}_{\bot}|\over E}\sim\theta_1\equiv {\mu \over E} \ll 1\>, 
\eeq
where $\Theta_s$ is the typical electron scattering angle. 

Let us consider the probability amplitude to scatter on $N$ static centres, and call the incoming and outgoing electron 4-momenta $p_0$ and $p_N$, 
respectively.
In the high energy limit where electron mass and spin effects are irrelevant, 
the $S$-matrix element corresponding to multiple 
          elastic 
scattering reads \cite{LL}
\beql{Sscatt1}
S_{scatt} \propto \delta (p_N^0 - p_0^0) \sum_{\sigma} \int \prod_{i=1}^{N-1}
 \left [
{d^3\vec{p}_i \over p_i^2 + i \eta} \right ] \prod_{i=1}^N 
\left [ {e^{-i \vec{q}_i
\cdot \vec{x}_{\sigma(i)}} \over \vec{q}_i^{\ 2} + \mu^2} \right ]  \>,
\eeq
where after the $i^{th}$ momentum transfer $\vq{i}$, the electron momentum is
\bminiG{q-def}
\vec{p}_i = \vec{p}_{i-1} + \vec{q}_i \quad ; \quad i = 1, \ldots N \>,
\eeeq
and
\beeq
p_i^0 = E  \quad ; \quad i = 0, \ldots N-1 \>.
\emini
The resulting cross section may be worked out in the eikonal limit of small 
angle scattering (see e.g. \cite{GW}). Let us indicate the main steps of 
the derivation in the well known Glauber picture.

$S_{scatt}$ contains the phase factor $\exp(i\varphi_{scatt})$ with 
\beql{PhiScatt1}\eqalign{
 \varphi_{scatt}\> &=\> - \sum_{i=1}^N \vec{q}_i \cdot \vec{x}_{\sigma(i)} 
= -  \sum_{i=1}^N
\vec{q}_{i\bot} \cdot \vec{x}_{\sigma(i)\bot} +  \sum_{i=1}^{N-1} p_{i||} 
(z_{\sigma(i+1)} - z_{\sigma(i)}) \cr
& +  \left ( p_{0||} \ z_{\sigma(1)} - p_{N||} \ z_{\sigma(N)} \right )\>,
}\eeq
where $z_{\sigma(i)}$ is the longitudinal component of $\vec{x}_{\sigma(i)}$ 
and we sum over all permutations $\sigma$ of \{$1, \ldots N$\}.

$S_{scatt}$ may be calculated by integrating successively over the 
longitudinal momenta $p_{i||}$, for $i = 1, \ldots N-1$. Let us start with 
$\int dp_{1||}$. 

When $(z_{\sigma(2)} - z_{\sigma(1)}) > 0$, the integration over $p_{1||}$ 
is performed by closing the integration contour in the upper half of the 
complex $p_{1||}$ plane (see (\ref{PhiScatt1})), and taking the pole 
$p_{1||} = E - \displaystyle{{p_{1 \bot}^2 / 2E}} + i\eta$.

When $(z_{\sigma(2)} - z_{\sigma(1)}) < 0$, we have to close the contour in 
the lower half-plane. 
We take the pole $p_{1||}= -E + \displaystyle{{p_{1 \bot}^2 / 2E}} -i\eta$,
but this yields a residue which is suppressed by a factor of order 
$\displaystyle{(p_{1 \bot}^2 + \mu^2) / E^2}$ when compared to the first 
case, where we have chosen $\vec{p}_{0 \bot} = 
0$. 

Finally, whatever the sign of $(z_{\sigma(2)} - z_{\sigma(1)})$ is, 
the poles in $p_{1||}$ corresponding to the "Coulomb denominator" 
$(\vec{q_1}^2 + \mu^2)$ lead to residues involving a suppression factor 
$\sim$ $\exp\{-\mu |z_{\sigma(2)} - z_{\sigma(1)}|\}$, 
and may be neglected when $\lambda \gg \mu^{-1}$. 
This may be repeated for all $p_{i||}$ integrations.

We deduce from the above~:
\begin{itemize}
\item
The high energy electron scatters successively on ordered scattering centres 
(in the $z$ direction)~:
$z_{\sigma(N)} > z_{\sigma(N-1)} > \ \ ...\ \ > z_{\sigma(1)}$. 
In (\ref{Sscatt1}), only the identity permutation has to be kept and backward 
scattering may be neglected\footnote{In other words the permutations 
different from identity imply the occurence of 
vacuum creation of $e^{+}e^{-}$ pairs, 
which process is 
power-suppressed when $E \to \infty$.}. 
\item
As a consequence  of (\ref{ind.scatt.}),
the internal electron momenta are on-shell, 
which corresponds to independent elastic scatterings on static centres:
\beq
p_{i||} \simeq E - {p_{i_{\bot}}^2 \over 2E}\>, \quad i = 1,2, \ldots N-1\>;
\qquad p_{0||}= E\>.
\eeq
\end{itemize}
The phase (\ref{PhiScatt1}) takes the form
\beql{PhiScatt2}
\varphi_{scatt}  \>\>=\>\>  \frac{p_{N\perp}^2}{2E}z_N\>-\> \sum_{i=1}^N
\vec{q}_{i\bot} \cdot \vec{x}_{i\bot} \>-\> \sum_{i=1}^{N-1} {p_{i_{\bot}}^2 
\over 2E}  (z_{i+1} - z_{i})\>.
\eeq
Using $\vec{q}_{i\perp}$ as integration variables we obtain,
up to an irrelevant phase and the energy conservation $\delta$-factor,
\beql{Sscatt2}
S_{scatt} \>\propto\> 
\int \prod_{i=1}^N \left [ d^2 \vec{q}_{i\bot} {e^{-i \vec{q}_{i\bot} 
\cdot \vec{x}_{i\bot}} \over 
\vec{q}_{i\bot}^{\ 2} + \mu^2} \right ] \cdot \prod_{i=1}^{N-1}  
\left [ e^{-i {p_{i\bot}^2 \over 2E}  (z_{i+1} - z_{i})} \right ]
\cdot \delta^2 \left ( \sum_{i=1}^N \vec{q}_{i\bot} -
\vec{p}_{N_{\bot}}\right ) .
\eeq
The scattering amplitude squared reads
\bea
|M_{scatt}|^2 \propto \int \prod_{i=1}^N \left [
d^2\vec{q}_{i_{\bot}} d^2 \vec{q}\,_{i_{\bot}}' \ 
{e^{i(\vec{q}\,_{i_{\bot}}' - \vec{q}_{i_{\bot}}) 
\cdot \vec{x}_{i_{\bot}}} \over (\vec{q}^{\ 2}_{i\bot} 
+ \mu^{2}) (\vec{q}\ '^{2}_{i\bot} + \mu^{2})} \right ] \prod_{i=1}^{N-1}
\left [ e\pw {i  {{p_{i\bot}^{'2} - p_{i\bot}^{2}}\over {2E}} (z_{i+1}-z_{i})}
 \right ] \nn \\
\cdot \delta ^2 \left ( \sum_{i=1}^N \vec{q}_{i_{\bot}} -
\vec{p}_{N_{\bot}} \right ) \delta^2 \left ( \sum_{i=1}^N 
\vec{q}\ '_{i_{\bot}} - \vec{p}_{N_{\bot}} \right )\>.  
\eea
The next step is to average over the coordinates $\vec{x}_{i}$. 
Since the centres are assumed to be uniformly distributed, averaging over
$\vec{x}_{i\perp}$ leads to $\vec{q}\,_{i\perp}'=\vec{q}\,_{i\perp}$, 
$\vec{p}\,_{i\perp}'=\vec{p}\,_{i\perp}$, 
which results in 
\beq
{d\sigma_{scatt}\over d^2 \vec{p}_{N\bot}} \propto \int  \prod_{i=1}^N  
{d^2 \vec{q}_{i_{\bot}}
\over (\vec{q}_{i_{\bot}}^{\ 2} + \mu^2)^2} \> \delta ^2 \left (
\sum_{i=1}^N \vec{q}_{i_{\bot}} - \vec{p}_{N_{\bot}} \right )\>.
\eeq
Here we have dropped an overall factor including the transverse surface 
$\delta^2(\vec{0})$. 
Integrating over $\vec{p}_{N\bot}$ we finally arrive at
\beql{SigmaScatt}
 \sigma_{scatt} \propto \int \prod_{i=1}^N {d^2 \vec{q}_{i_{\bot}}
\over (\vec{q}_{i_{\bot}}^{\ 2} + \mu^2)^2} \>.
\eeq
A similar expression may be found in \citd{Levin}{LR}.

\mysection{Radiation spectrum induced by multiple  scattering in the soft 
photon approximation}
\label{sec:nom3}
Multiple scattering induces radiation. The calculation of the \br probability 
is done in the soft photon approximation
\beql{softapp}
\omega\ \ll \ E \>.
\eeq
This allows 
to factorize the radiation amplitude as the product of the 
multiple scattering amplitude times the photon emission amplitude. 
The ratio between the radiation cross section and the  scattering cross 
section  yields the radiation probability.

Let us consider the probability amplitude for emitting a photon of 4-momentum 
$k=(\omega,\vec{k}\,)$ 
off the electron between centres $\vec{x}_j$ and $\vec{x}_{j+1}$. 
The 
   corresponding
$S$-matrix element reads
\beal{Sjrad}
S_{rad}^j \propto e \delta (p_N^0 - p_0^0) \int \prod_{i=1}^{j-1} 
\left [ {d^3 \vec{p}_i
\over p_i^2 + i \varepsilon} \right ] \cdot \int d^3 \vec{p}_j 
{\varepsilon \cdot p_j \over k \cdot p_j} \left [ {1 \over (p_j - k)^2 + i
\eta} - {1 \over p_j^2 + i \eta} \right  ] \nn \\
\cdot \int \prod_{i=j+1}^{N-1} \left [ {d^3 \vec{p}_i \over (p_i - k)^2 
+ i \eta }
\right ] \cdot \prod_{i=1}^N \left [ {e^{-i\vec{q}_i \cdot \vec{x}_i} 
\over \vec{q}_i^{\ 2} + \mu^2}
\right ]\>, 
\eea
with 
the momentum variables $p_i$ defined in (\ref{q-def}).
The two physical photon polarization states 
$\varepsilon$
are chosen in the form 
\beql{physpol}
\varepsilon = ( \varepsilon_0 , - \varepsilon_0 , \vec{\varepsilon}_{\bot} ) 
\quad ; \qquad \varepsilon\cdot k=0 \>\Longrightarrow\>\> 
\varepsilon_0 = \frac{\vec{\varepsilon}_{\bot} \cdot \vec{k}_{\bot}}
{\omega + k_{||}} \simeq \frac{\vec{\varepsilon}_{\bot} \cdot 
\vec{k}_{\bot}}{2\omega} \>.
\eeq
In the same way as in the previous section, the phase $\varphi_{rad}$ 
is given by the r.h.s. of (\ref{PhiScatt1}) and the integrations over 
$p_{i||}$ are done by closing the contour in the upper half-plane,
\bminiG{poles}\label{pole1}
p_i^2 = 0 \ &\Rightarrow& \   p_{i||} \simeq E - {p_{i_{\bot}}^2 \over 2E}\>, \\
\label{pole2}
(p_i - k)^2 = 0 \ &\Rightarrow& \   p_{i||} - k_{||} \simeq E - \omega 
- {(p_i - k)^2_{\bot} \over 2(E - \omega)} \>.
\emini
As far as the intermediate state $j$ is concerned, we may have
either $(p_j-k)^2=0$ or  $p_j^2=0$.
Since the accompanying radiation factor is invariant,
\beq
   \frac{\varepsilon \cdot p_j}{ k \cdot p_j}\>=\> 
   \frac{\varepsilon \cdot (p_j-k)}{ k \cdot (p_j-k)}\>; 
\qquad \left[\, \varepsilon\cdot k=0\>, \>\> k^2=0\>\right],
\eeq
it may be always expressed in terms of the {\em real}\/ electron momentum.
For example, taking $p_j^2=0$ we obtain
\bminiG{numden}
\varepsilon \cdot p_j &=&\varepsilon_0(E+p_{j||})
- \vec{\varepsilon}_\perp\cdot \vec{p}_{j\perp}\>\simeq\> 
E\,\vec{\varepsilon}_\perp\!\cdot \left(\frac{\vec{k}_\perp}{\omega}
-\frac{\vec{p}_{j\perp}}{E}\right) \>\equiv\>E\,\vec{\varepsilon}_\perp\!\cdot 
\vec{u}_{j} \>, \\
k \cdot p_j &=&  {E \omega \over 2} 
\left ( {\vec{k}_{\bot} \over \omega} -
{\vec{p}_{j_{\bot}} \over E} \right )^2 \>\equiv\> \frac{E\omega}{2} u_j^2\>.
\emini 
Here we introduced the dimensionless transverse vector $\vec{u}_j$ which, 
in the quasi-collinear kinematics under interest, measures the angle   
between the photon and the electron after scattering number $j$,
\beql{uj}
 \vec{u}_j\>=\> \frac{\vec{k}_\perp}{\omega}-\frac{\vec{p}_{j\perp}}{E}\>=\> 
\frac{\vec{k}_{\bot}}{\omega} - \sum_{i=1}^j \frac{\vec{q}_{i_{\bot}}}{E} \>.
\eeq
For another pole, $(p_j-k)^2=0$, one has to substitute in (\ref{numden})
\beq
\vec{u}_j 
\>\> \Longrightarrow 
\>\>  \frac{\vec{k}_\perp}{\omega}-\>\frac{\vec{p}_{j\perp}-\vec{k}_\perp}{E}
\>=\>\vec{u}_j\left\{1\>+\>\cO{\frac{\omega}{E}}\right\}\>\simeq\> \vec{u}_j\>,
\eeq
which difference, however, may be neglected in the soft approximation 
(\ref{softapp}). 

Thus, we can write the $S$-matrix element (\ref{Sjrad}) as
\beql{Sraddiff}
S_{rad}^j \propto e \delta (p_N^0 - p_0^0) \int \prod_{i=1}^{N-1} \left [ d^2
\vec{p}_{i_{\bot}} \right ] \prod_{i=1}^N 
\left [ {1 \over \vec{q}_{i_{\bot}}^{\ 2} + \mu^2}
\right ] \cdot {\varepsilon \cdot p_j \over k \cdot p_j} \left ( \left . e^{i
\varphi_{rad}} \right |_{(p_j - k)^2=0} - \left . e^{i \varphi_{rad}} 
\right |_{p_j^2 = 0}
\right )\>,
\eeq
where the elementary soft radiation factor is
\beql{residu}
\left . {\varepsilon \cdot p_j \over k \cdot p_j} \right |_{(p_j - k)^2=0} 
\simeq \left .
{\varepsilon \cdot p_j \over k \cdot p_j} \right |_{p_j^2=0} 
= {2 \over \omega} \ 
\vec{\varepsilon}_{\bot} \cdot {\vec{u}_j \over u_j^2} \>,
\eeq 
with $\vec{u}_j$ defined in (\ref{uj}).

Now we must express the phases $\varphi_{rad}$:
\beq\eqalign{
 \varphi_{rad} &\>=\> -\sum_{i=1}^N \vec{q}_i\cdot\vec{x}_i
= -\sum_{i=1}^N\vec{q}_{i\perp}\cdot\vec{x}_{i\perp} \>+\> \varphi_{||} \>, \cr
\varphi_{||} &\>=\>\sum_{i=1}^{N} {q}_{i||}{z}_{i}
\>=\> p_{0||}z_1-p_{N||}z_N 
+ \sum_{i=1}^{N-1} p_{i||}(z_{i+1}-z_i) \>. 
}\eeq
According to (\ref{poles}), the value of the longitudinal component 
$p_{i||}$ depends on whether $p_i^2=0$ or $(p_i-k)^2=0$. 
Expanding (\ref{pole2}) in $\omega/E$ and neglecting $\cO{\omega^2/E}$ for 
the second case, we derive
\beql{pole2mod}
 p_{i||} \>=\> \left(E-\frac{p_{i\perp}^2}{2E}\right) 
 \>-\>\frac{\omega}{2}\,u_i^2\>, \qquad\mbox{for}\>\> (p_i-k)^2=0\,.
\eeq 
Using (\ref{pole1}) and (\ref{pole2mod}) and invoking 
(\ref{PhiScatt2}) for the scattering phase, we represent the phase of the
first term in (\ref{Sraddiff}), corresponding to $(p_j - k)^2 = 0$, as
\beql{PhiRad-PhiScatt1}\eqalign{
\varphi_{rad} &\>=\> \varphi_{scatt}\>+\> \Phi_j\>, \cr
\Phi_j &\>=\> - {\omega \over 2} \sum_{i=j}^{N-1} u_i^2 (z_{i+1} - z_i) \>.
}\eeq
Similarly, for $p_j^2 = 0$, we use (\ref{pole1}) for $i = 1, \ldots j$ and 
(\ref{pole2mod}) for $i = j+1, \ldots N-1$ to obtain the expression for 
$\varphi_{rad}$, which differs from (\ref{PhiRad-PhiScatt1}) 
simply by substituting $j+1$ for $j$, that is $\Phi_j \to \Phi_{j+1}$
in (\ref{PhiRad-PhiScatt1}).

We stress that the phase difference depends only on the longitudinal 
separations. 

Using the picture of the electron moving along a classical trajectory between 
scattering centres, 
$\vec{x}_{i+1} - \vec{x}_i = \vec{v}_i \  (t_{i+1} - t_i)$ 
with $\vec{v}_i$ the electron velocity, 
one may write the phase in Lorentz-invariant form as
\beql{classical}
\Phi_j \>\simeq\>  k \cdot (x_j - x_N) \>.
\eeq

After summing $S^{j}_{rad}$ over $j$ and adding the radiation off initial 
and final lines we get the total radiation $S$-matrix element
\bminiG{Srad}
S_{rad} \>\propto\>  e \delta (p_N^0 - p_0^0) \int \prod_{i=1}^N \left [ 
\frac{d^2\vec{q}_{i\bot}}{\vec{q}_{i\bot}^{\ 2} + \mu^2} \right]
 \cdot  e^{i\varphi_{scatt}} 
\cdot \delta^2 \left ( \sum_{i=1}^N \vec{q}_{i\bot} -
\vec{p}_{N_{\bot}}\right ) \cdot {\cal{R}} \>,
\eeeq
where
\beeq
{\cal{R}} = \sum_{j=1}^{N-1} {\varepsilon \cdot p_j \over k \cdot p_j} 
\left ( e^{i\Phi_j} - e^{i\Phi_{j+1}} \right ) 
- \frac{\varepsilon \cdot p_0}{k\cdot p_0} e^{i\Phi_1} 
+ \frac{\varepsilon \cdot p_N}{k\cdot p_N} 
=  \sum_{j=1}^N \left( {\varepsilon \cdot p_j \over k \cdot p_j} 
- {\varepsilon \cdot p_{j-1}\over k\cdot p_{j-1}} \right ) e^{i\Phi_j} \>.
\emini
Thus,
the integrand of $S_{rad}$ is proportional to that of $S_{scatt}$ 
in (\ref{Sscatt2}). 
As  seen from (\ref{PhiRad-PhiScatt1}), 
the proportionality factor $e{\cal{R}}$ 
does not depend on the transverse coordinates $\vec{x}_{i\bot}$. 
Therefore, after squaring (\ref{Srad}) to get $|M_{rad}|^2$, 
we may follow the same averaging procedure over $\vec{x}_{i\bot}$ 
as for $|M_{scatt}|^2$. 
The result is
\beq
|M_{rad}|^2 \propto \ e^2\ \int  \prod_{i=1}^N  {d^2 \vec{q}_{i_{\bot}}
\over (\vec{q}_{i_{\bot}}^{\ 2} + \mu^2)^2}  \delta ^2 \left (
\sum_{i=1}^N \vec{q}_{i_{\bot}} - \vec{p}_{N_{\bot}} \right )
\sum_{pol} \left | {\cal{R}} \right |^2 
\>.
\eeq 
Integrating over $\vec{p}_{N\bot}$ gives
\beq
d\sigma_{rad} \propto \ e^2\ \left [ \int  \prod_{i=1}^N  
{d^2 \vec{q}_{i_{\bot}}
\over (\vec{q}_{i_{\bot}}^{\ 2} + \mu^2)^2} \sum_{pol} \left | 
{\cal{R}} \right |^2 \right ] {d^3 \vec{k} \over (2 \pi )^3 2 \omega}\>.
\eeq
Finally, 
using (\ref{residu}) and summing over the photon polarizations leads 
to the radiation intensity 
\bminiG{Spectrum}
\omega {dI \over d \omega} = {\alpha \over \pi^2} \int d \Omega \left < 
\left | \sum_{i=1}^N \vec{A}_i \ e^{i\Phi_i} \right |^2 \right >\>,
\eeeq
with $d\Omega$ representing the integration over the photon direction.
Here we have defined the emission current
\beeq\label{current}
\vec{A}_i = \displaystyle{{\vec{u}_i \over u_i^2}} -
\displaystyle{{\vec{u}_{i-1} \over u_{i-1}^2}} \>.
\emini
The brackets denote the averaging over transverse momenta and longitudinal 
coordinates using normalized probability distributions,
\beql{averaging}
\left < \ (\ \ ...\ \ )\ \right >\  \Longleftrightarrow \ 
\int \prod_{\ell=1}^{N-1} \frac{d\Delta_\ell}{\lambda}
 \exp\left(-\frac{\Delta_\ell}{\lambda}\right) \cdot
\int  \prod_{i=1}^N 
 {\mu^2d^2 \vec{q}_{i_{\bot}}
\over \pi (\vec{q}_{i_{\bot}}^{\ 2} + \mu^2)^2}\>\>
(\ \ ...\ \ )\ \>,
\eeq
where the exponential factor reflects the survival probability 
of the electron over the distance $\Delta_{\ell} = z_{\ell +1} - z_{\ell}$. 

In a general case of scattering potential other than Coulomb, 
the averaging over transverse momenta has the form
\beql{normXsec}
 \int  \prod_{i=1}^N d^2 \vec{q}_{i_{\bot}} \> V(q_{i\perp}^2)
\>(\ \ldots\ )\>; \qquad \int d^2\vec{q}\>\> V(q^2)=1\>,
\eeq
with $V(q^2)$ the normalized cross section for elastic electron 
scattering in the medium. 

For the sake of completeness and also in view of the generalization to 
nonabelian radiation processes, it is useful to mention the old-fashioned 
perturbation theory proof of (\ref{Spectrum}).

Let us consider the radiation amplitude induced by $N$ transfers of fixed 
momenta at centres $\vec{x}_i$ and times $t_i$. In time-ordered perturbation 
theory, in the limit $E \to \infty$, $M_{rad}^j$ is obtained by integrating 
the phase factor over the emission time $\tau_j$, lying between $t_j$ and 
$t_{j+1}$, 
\beql{TOPT1}
M_{rad}^j = - i e {2
\varepsilon \cdot p_j \over 2E} \left \{ \int_{t_j}^{t_{j+1}} d \tau_j 
\ e^{i \tau_j (|\vec{p}_j -
\vec{k}| + \omega - |\vec{p}_j|)} \right \} \cdot e^{i \delta \varphi}
M_{scatt} \>.
\eeq
We denote by $\delta \varphi$ the phase difference between $M_{rad}^j$ and 
$M_{scatt}$. For the interaction times $t_i$, $i > j$, the associated phase 
is $t_i(|\vec{p}_i| -
|\vec{p}_{i-1}|) \simeq 0$ for $M_{scatt}$,  but becomes $t_i(|\vec{p}_i 
- \vec{k}| -
|\vec{p}_{i-1} - \vec{k}|) \simeq t_i \ \vec{k} \cdot (\vec{v}_{i-1} 
- \vec{v}_i)$
for $M_{rad}^j$. Thus
\beql{TOPT2}
\delta \varphi = \sum_{i=j+1}^N t_i \ \vec{k} \cdot \left ( \vec{v}_{i-1} -
\vec{v}_i \right ) \>.
\eeq
Using
\beq
\vec{x}_{i+1} - \vec{x}_i = \vec{v}_i \left ( t_{i+1} - t_i \right ) \>, 
\eeq
and
\beql{TOPT3}
| \vec{p}_i - \vec{k} | - |\vec{p}_i| +
\omega \simeq \omega - \vec{k} \cdot \vec{v}_i = {k \cdot p_i \over E} \>,
\eeq
we easily get
\beql{TOPT4}
M_{rad}^j =
\ e \ {\varepsilon \cdot p_j \over k \cdot p_j} \left ( e^{ik\cdot x_j} 
- e^{ik \cdot x_{j+1}}
\right ) M_{scatt} \>.
\eeq
Integrating over $\tau_j$ in (\ref{TOPT1}) gives two contributions, 
proportional to $e^{ik \cdot x_j}$ and $e^{ik \cdot x_{j+1}}$, which may be 
respectively included in radiation amplitudes induced by the $j^{th}$ and 
$(j+1)^{th}$ scatterings. Finally, the radiation amplitude induced by $N$ 
scatterings is found to be identical to the previous result (\ref{Spectrum}).

\mysection{Explicit calculation of the radiation intensity}
\label{sec:nom4}
Here we present the analytic calculation of the energy spectrum 
(\ref{Spectrum}) in the soft photon approximation.
The spectrum (\ref{Spectrum}) may be written in two equivalent forms 
as follows:
\bminiG{SpectrumFactLPMBH}
\label{SPBH} 
\omega {dI \over d \omega} &=& {\alpha \over \pi^2} \int d\Omega \ 
\left < 2\Re \sum_{i=1}^N \sum_{j=i+1}^N 
\vec{A}_i \cdot \vec{A}_j \> 
e^{i\Phi_{ji}}
 + \sum_{i=1}^N\left|  \vec{A}_i \right |^2 \right > \\
\label{SPFA}
&=& {\alpha \over \pi^2} \int d\Omega \ \left < 
2\Re \sum_{i=1}^N
\sum_{j=i+1}^N \vec{A}_i \cdot \vec{A}_j \left ( 
e^{i\Phi_{ji}}- 1 \right ) + \left |
\sum_{i=1}^N \vec{A}_i \right |^2 \right >\>,
\emini
with the relative phase
\beq
  \Phi_{ji}\>=\> \Phi_j-\Phi_i \>=\> \frac{\omega}{2}
\sum_{\ell=i}^{j-1} u_\ell^2 (z_{\ell+1}-z_{\ell}) \>.
\eeq

\subsection{Heuristic derivation of the LPM spectrum}
The radiation pattern depends crucially on the phases. 
If $\Phi_{ji}$ are large, the rapid oscillations in the first term of
(\ref{SPBH}) wash this contribution away, and we have the Bethe-Heitler
regime of independent radiation.   
Dividing by the size of the medium, $N \lambda$, 
for the differential energy spectrum {\em per unit length}\/ 
(the radiation density) we obtain  
\beql{BH}\eqalign{
\omega {dI \over d \omega dz} &= \frac1{\lambda}\> {\alpha \over \pi^2} 
\int d \Omega  \> \>\frac1N 
\left < \sum_{i=1}^N \left |\vec{A}_i  \right |^2 \right>
\>=\> {\alpha \over\pi\lambda} \int \frac{d^2\vec{u}_1}{\pi}\>
\left\langle\left |\vec{A}_1  \right |^2 \right\rangle \cr
&= {\alpha \over\pi\lambda} \int \frac{d^2\vec{u}}{\pi}\>
\lrang{\frac{\Theta_s^2}{u^2(\vec{u}-\vec{\Theta}_s)^2}}
\>\simeq\> \frac{2\alpha}{\pi\lambda}\lrang{\ln\frac{\Theta_s^2}{u_{min}^2}}\>,
}\eeq  
with $u_{min}^2\ll\Theta_s^2$ 
the value of the emission angle below which the phase
between neighbouring centres becomes small: 
\beq
  \Phi_{i,i+1}= \half{\omega}\,u_{i}^2(z_{i+1}-z_i)
\>\sim\> \half{\omega\lambda}\,u_{i}^2 \> < \>1\> \Leftrightarrow u_{i}^2 \> 
< \>u_{min}^2\>; 
\quad u_{min}^{-2}= \half{\omega\lambda}\,.  
\eeq 
Small emission angles $u^2<u_{min}^2$ or $(\vec{u}-\vec{\Theta_s})^2<u_{min}^2$
correspond to photon formation time {\em larger}\/ than the distance 
to the neighbouring centres,
\beq
 t_{form} \>=\>\frac{2\omega}{k_{\perp}^2} =\frac2{\omega u^2}\> > \>\lambda\,;
\eeq
so that destructive interference with the radiation due to the nearest neighbour 
screens the logarithmic divergence.
Introducing the characteristic dimensionless parameter $\kappa$, representing 
the typical phase difference between neighbouring centres,
\beql{kappa}
\kappa = {\lambda \mu^2 \over 2} {\omega \over E^2} \>,
\eeq
we may obtain (\ref{BH}) as 
\beql{BH2}
 \left(\omega {dI \over d \omega dz} \right)^{(\mbox{\scriptsize BH})}
\>=\> 
\frac{2\alpha}{\pi\lambda}\lrang{\ln\frac{\kappa\,q^2}{\mu^2}} 
\>\simeq\>\frac{2\alpha}{\pi\lambda}\> \ln\kappa\>. 
\eeq 
The Bethe-Heitler limit corresponds to $\kappa>1$.

In the opposite limit, $\kappa\ll1$, the phases between neighbouring centres 
are vanishingly small, so that a group of $\nu$ centres radiates coherently. 
To estimate the {\it coherence number} $\nu$~\cite{GGS}, we look for 
the separation between two centres, 
such that an accumulated phase becomes of order unity.
In a first approximation one may assume that the trajectory of the electron is 
a random walk in the transverse momentum space,
\beq
{u_{\ell +1}^2} \simeq  {u_{\ell}^2} + \lrang{\Theta_s^2} \>.
\eeq
The phase $\Phi_{i,i+\nu}$ is estimated then 
as 
\beql{phase}
\Phi \ \approx \ \half\omega\lambda
\left[\nu u_{i}^2 +\half \nu(\nu-1)\lrang{\Theta_s^2}\right ]
\>\approx\> \kappa \left[\nu \cdot \frac{u_i^2}{\theta_1^2} + 
\frac{\nu^2}{2} \cdot \frac{\lrang{\Theta_s^2}}{\theta_1^2}\right]  \> \sim\>1 \>.
\eeq
As will become apparent later from (\ref{ang.constraint1}), the second term here 
is $\nu$ times bigger than the first one, which gives
\beql{estnu}
\nu^{-1} \>\approx\> \sqrt{\frac{\kappa}2\,\frac{\lrang{\Theta_s^2}}{\theta_1^2}}
 \>\ll\>1\>.
\eeq
If the {\it coherence length} exceeds the size of the medium, $\nu \lambda > L$, 
all the phases are small, and the first term in (\ref{SPFA}) vanishes. 
In the last term the emission currents (\ref{current}) 
add up into the expression
\beq
\sum_{i=1}^N \vec{A}_i = {\vec{u}_N \over u_N^2} - {\vec{u}_0 \over u_0^2} 
\quad ; \quad
\vec{u}_N = \vec{u}_0 - \frac{\vec{q}_{tot}}{E} \>,
\eeq
which corresponds to the radiation induced by a single scattering with 
the momentum transfer 
$\vec{q}_{\perp tot} = \displaystyle{\sum_{i=1}^N} \vec{q}_i$.
It is usually called the {\em factorization}\/ limit.
For a given $\vec{q}_{tot}$ the radiation is independent of the size 
and the properties of the medium and reads
\beq
\omega {dI \over d \omega} = {2 \alpha \over \pi} \ \left < \ln 
{q_{tot}^2 \over m^2} \right > \>.
\eeq
Here we have introduced a finite electron mass $m$ to regularize 
the collinear divergence. 
It is the only place where the electron mass enters in the high-energy limit. 
Otherwise, the induced radiation is collinear-safe. 
When the size of the medium is large enough to embody a few coherence lengths,
$L \gg \nu\lambda$, the LPM-suppression of the Bethe-Heitler spectrum takes 
place. 
In order to quantify it, it suffices to ``slice'' the medium and substitute 
the number of ``effective radiators'' $N/\nu$ for $N$.  
For the radiation density we write 
\beql{LPM}
\omega {dI \over d\omega dz} \approx {1 \over \nu}
\left(\omega {dI \over d\omega dz}\right)^{\mbox{\scriptsize (BH)}}
\>.
\eeq

In the standard Bethe-Heitler spectrum off a point-like source (\ref{BH}) 
the logarithmic enhancement factor appears due to emission angles much smaller
than the scattering angle. The effective centre, however, radiates at typical 
angles $u^2$ such that the formation time $2 (\omega u^2)^{-1}$ is 
of the order of the length of the radiator $\nu \lambda$. 
This implies an emission angle $u^2$ of the order of the accumulated 
diffusion angle,
\beql{ukest}
u^2\>\sim\> \frac{1}{2}\nu\lrang{\Theta_s^2}\>.
\eeq
Under these conditions the logarithmic enhancement is absent, and (\ref{LPM}) 
becomes
\beql{LPMest}
\omega {dI \over d\omega dz} \approx {1 \over \nu}
\cdot \frac{2\alpha}{\pi\lambda}\,C
\>=\> C\, \frac{\alpha}{\pi\lambda}\>
 \sqrt{2{\kappa}\,\frac{\lrang{q_\perp^2}}{\mu^2}}  \>=\>
C\, \frac{\alpha}{\pi}\sqrt{\,\omega\,\frac{\lrang{q_\perp^2}}{\lambda E^2}}
\>=\>C\, \frac{\alpha}{\pi}\sqrt{\,\omega\,\frac{\lrang{\Theta_s^2}}{\lambda}} \>,
\eeq
with $C$ a constant of order 1.
This reproduces the well-known Landau-Pomeranchuk-Migdal result ($C=1$).
Strictly speaking, the above derivation based on the random walk picture
applies only to the cases when the mean squared momentum transfer in a single
scattering is well defined.
However, this is not true for Coulomb scattering where the integral 
determining $\lrang{q_\perp^2}$ formally diverges logarithmically:
\beq
 \lrang{q_\perp^2} = \int\frac{\mu^2\,dq^2}{(\mu^2+q^2)^2}\cdot q^2
= \mu^2 \int\frac{\theta^2\,d\theta^2}{(\theta_1^2+\theta^2)^2}=
\mu^2\,\ln \frac{\Theta^2_{max}}{\theta_1^2}\>.
\eeq 
Replacing the upper limit of the angular integral by the characteristic 
angle of the problem $u^2$ from (\ref{ukest}), 
we obtain 
\beq
  \frac{\lrang{q_\perp^2}}{\mu^2} = \frac{\lrang{\Theta_s^2}}{\theta_1^2}
 \>=\> \ln \left( \frac1{\sqrt{\kappa}}\,
       \sqrt{\frac{\lrang{\Theta_s^2}}{\theta_1^2}} \right)
 \>\simeq\> \half \ln\frac1\kappa \>,
\eeq
where we have neglected the log-factor under logarithm.
Substituting this ratio into (\ref{LPMest}), for the Coulomb case
we derive
\beql{LPMClmbest}
\left( \omega {dI \over d\omega dz}\right)_{\mbox{\scriptsize Coulomb}}
\>\simeq\> \frac{\alpha}{\pi\lambda}\>\sqrt{\kappa\ln\frac1\kappa}\,.
\eeq
This heuristic estimate coincides with the true answer derived below.

The origin of the extra logarithmic enhancement in (\ref{LPMClmbest})
is readily understood. The singularity of the Coulomb potential at small
distances corresponds to a long tail in the transverse
momentum distribution. This enriches the contribution from
``large jumps'', when the momentum transfer exceeds the inverse
Debye radius, $q_\perp^2\gg\mu^2$. 
Effective ``random walk'' steps become (logarithmically) larger, which reduces 
by the $\sqrt{\ln\kappa^{-1}}$ factor the coherence number $\nu$ and, thus,
the LPM suppression. 

Hereafter we shall concentrate on the coherent LPM regime  
\bminiG{LPMregime}\label{LPMregime1}
1\ \ll \nu \ \ll \ N \quad\Longrightarrow\quad 
 \frac{1}{N^2} \ \ll \ \kappa \ \ll \ 1 \>.
\eeeq
In terms of the photon energy the coherent region is limited by
\beeq\label{LPMregime2}
\frac{L_{cr}^2}{L^2}\equiv\frac{\lambda E}{L^2 \mu^2}
\>\><\>\> \frac{\omega}{E} \>\><\>\>
\frac{2E}{\lambda \mu^2}\equiv\frac{E}{E_{\mbox{\scriptsize LPM}}} \>.
\emini

\subsection{General expression for the induced radiation spectrum}
We turn to the original expression for the radiation spectrum 
(\ref{SpectrumFactLPMBH}).
As we shall see below, a finite (though large) number of scatterings 
($j-i\sim \nu$) is essential, so that for $N \gg \nu$
a given term of the double sum clearly depends only on the relative 
position of the two centres, $j-i=n+1$, $n\ge0$.
Moreover, the internal sum over $j$
in (\ref{SPFA})
converges and, therefore, does not depend on $i$ in the same approximation.
The sum over $i$ then gives the total number of scatterings $N$. 

Dividing by the size of the medium, $N \lambda$, 
and neglecting the factorization contribution in the $N\to\infty$ limit,
we obtain the following expression for 
the differential energy spectrum per unit length:
\beal{U-SpectrumFactLPMBH} 
\omega {dI \over d \omega dz} = 
{\alpha \over \pi\lambda} 
\int {d^2 \vec{U}_1 \over \pi} \ \left < 2\Re \sum_{n=0}^{\infty}
 \vec{J}_1  \cdot 
\vec{J}_{n+2} \left [ \exp \left \{ i \kappa \sum_{\ell = 1}^{n+1} U_{\ell}^2 
{{z_{\ell + 1} -z_{\ell}} \over \lambda} \right \} - 1 \right] 
\right > \left(1 + \cO{\frac{1}{N}}\right).
\eea
Here we have introduced new variables $\vec{U}_{\ell}$ to represent
$\vec{u}_{\ell}$ in units of the typical scattering angle $\theta_1=\mu /E$~:
\beql{U}
\vec{U}_{\ell} = \theta_1^{-1}\cdot \vec{u}_{\ell} \>;
\qquad d\Omega= \theta_1^2 \>d^2\vec{U}\>,
\eeq
and rescaled the emission currents correspondingly:
\beql{JofU}
\vec{J}_i = \theta_1\cdot \vec{A}_i\>=\> {\vec{U}_i
\over U_i^2} - {\vec{U}_{i-1} \over U_{i-1}^2} \>.
\eeq
It is also convenient to express the transferred momenta in units of $\mu$,
\beql{Q}
\vec{Q}_{\ell} = \frac{\vec{q}_{\ell}}{\mu} \>, \qquad 
\vec{U}_\ell = \vec{U}_{\ell-1} -\vec{Q}_{\ell} \>,
\eeq
so that
$$
\abs{\vec{Q}_\ell}=  \abs{\vec{U}_{\ell}-\vec{U}_{\ell-1}}\sim 1\>.
$$
Performing in (\ref{U-SpectrumFactLPMBH}) 
the averaging over longitudinal separations 
with use of (\ref{averaging}), we obtain
\bminiG{exactLPM}
\omega {dI \over d\omega dz} =  {\alpha \over \pi\lambda} 
\int {d^2\vec{U}
\over \pi} \>2\Re \sum_{n=0}^{\infty} \int {\prod_{\ell = 1}^{n+2} 
d^2\vec{Q}_{\ell}\> V(Q_{\ell}^2) \>
\vec{J}_1 \cdot \vec{J}_{n+2} \left [ \prod_{m = 1}^{n+1} \psi (U_{m}^2) - 1 
\right ]} \>,
\eeeq
where
\beeq\label{psi}
\psi(U^2) = (1 - i \kappa U^2)^{-1} \>.
\emini
This holds in general for arbitrary interactions. For the particular case of 
Coulomb scattering, $V$ is given by
\beql{V}
V(Q_{\ell}^2)  = {1 \over \pi (Q_{\ell}^2 + 1)^2} \>.
\eeq

In (\ref{exactLPM}), the dependence on $\vec{Q}_1$ and $\vec{Q}_{n+2}$ is 
contained only in the product of currents
\beq
\vec{J}_1 \cdot \vec{J}_{n+2} = \left ( {\vec{U}_1 \over U_1^2} - {\vec{U}_1
+ \vec{Q}_1 \over (\vec{U}_1 + \vec{Q}_1)^2} \right ) \cdot \left (
{\vec{U}_{n+1} - \vec{Q}_{n+2} \over (\vec{U}_{n+1} - \vec{Q}_{n+2})^2} 
- {\vec{U}_{n+1} \over U_{n+1}^2}
\right ) \>.
\eeq
Keeping $\vec{U}_1,\, \vec{U}_2,\ldots \vec{U}_{n+1}$ fixed, 
we integrate first over $\vec{Q}_1$ 
(which is equivalent to integrating over $\vec{U}_0$, 
the direction of the incoming electron with respect to the photon) 
and $\vec{Q}_{n+2}$ (outgoing electron). 
To this end we define
\bminiG{ang.constraints}\label{ang.constraint1}
\vec{f}_0(\vec{U}_1) \equiv \int d^2 \vec{Q}_1 V(Q_1^2)\> \vec{J}_1 =
\pi{\vec{U}_1 \over U_1^2} \int dQ_1^2 \ V(Q_1^2) \ \Theta (Q_1^2 -
U_1^2) = \pi{\vec{U}_1 \over U_1^2} \int_{U_1^2}^\infty dQ^2\> V(Q^2) \>.
\eeeq
In the same way we have 
\beeq\label{ang.constraint2}
\vec{f}_0(\vec{U}_{n+1}) \equiv - \int d^2 \vec{Q}_{n+2} V(Q_{n+2}^2)
\vec{J}_{n+2} = 
 \pi{\vec{U}_{n+1} \over U_{n+1}^2}\int_{U_{n+1}^2}^\infty dQ^2\> V(Q^2) \>.
\emini
The spectrum can be rewritten as
\beql{U-spectrum}
\left . \omega {dI \over d \omega \ dz} =  {2 \alpha \over\pi\lambda}  
\Re \int \frac{d^2\vec{U}_1}{\pi}\>\vec{f}_0(\vec{U}_1)\cdot\vec{f}(\vec{U}_1)
\right |_{\kappa}^{\kappa=0} \>,
\eeq
where $\vec{f}(\vec{U}_1)$ is given by
\beq
\vec{f}(\vec{U}_1) = \psi (U_1^2) \vec{f}_0(\vec{U}_1) + \psi (U_1^2) 
\sum_{n=1}^{\infty} \prod_{\ell = 2}^{n+1}
\left [ \int d^2 \vec{Q}_{\ell} \ V(Q_{\ell}^2) \ \psi (U_{\ell}^2)
\right ] \vec{f}_0(\vec{U}_{n+1}) \>.
\eeq
The $\kappa$-dependence of $\vec{f}$ comes from the $\kappa$-dependence of 
$\psi$ (see (\ref{psi})).

\paragraph{Master equation.}
By expanding the sum over $n$, it is straightforward to see that 
$\vec{f}(\vec{U})$ satisfies the following integral equation
\beql{IntegralEquation}
(1 - i \kappa U^2) \vec{f} (\vec{U}) = \vec{f}_0(\vec{U}) + \int d^2 \vec{Q}
\ V(Q^2) \ \vec{f}(\vec{U} - \vec{Q}) \>.
\eeq
For $\kappa=0$ it has a trivial potential-independent solution
\beql{kapzersol}
 \vec{f}\,(\vec{U}\,) \>=\> \frac{\vec{U}}{U^2}\>.
\eeq
Indeed, substituting (\ref{kapzersol}) under the integral, we have
$$
 \int  d^2 \vec{Q}\ V(Q^2) \> \frac{\vec{U} - \vec{Q}}{(\vec{U} - \vec{Q}\,)^2}
\>=\> \pi\frac{\vec{U}}{U^2}\int_0^{U^2} dQ^2\> V(Q^2)\>.
$$ 
Taken together with (\ref{ang.constraints}) for $\vec{f}_0$, this gives 
for the r.h.s.
$$
 \pi\frac{\vec{U}}{U^2} \left\{ \int_{U^2}^\infty dQ^2\> V(Q^2) 
\>+\> \int_0^{U^2} dQ^2\> V(Q^2) \right\} \>=\> \frac{\vec{U}}{U^2}\cdot 1
$$
(due to the normalization of the scattering cross section (\ref{normXsec})),
which is identical to the l.h.s. ($\kappa=0$). 

In order to solve the equation (\ref{IntegralEquation}) for $\kappa\neq0$ 
we adopt a method different from the derivation advocated in \cite{BDPS}, 
which relied on an incorrect approximation. 
Going to the impact parameter space by defining the Fourier transform
\bmini
\widetilde{\vec{f}\ }(\vec{B}) &=&\int d^2\vec{U} \ e^{-i\vec{B}\cdot \vec{U}}
\vec{f}(\vec{U}) \>, \\
\widetilde{V}(B^2) &=& \int d^2\vec{Q}\  
e^{-i\vec{B}\cdot\vec{Q}}\  V(Q^2) \>,
\emini
we first derive the $B$-space image of the function (\ref{ang.constraints}):
\beql{fzero} 
\widetilde{\vec{f}_0}(\vec{B}) \>=\> - 2\pi i \ {\vec{B} \over B^2}\  (1 -
\widetilde{V}(B^2)) \>.
\eeq
In these terms (\ref{IntegralEquation}) converts into 
the differential equation
\beql{EDf}
(1 + i \kappa \vec{\nabla}_B^2)\  \widetilde{\vec{f_{\ }}}(\vec{B})\  =\ 
\widetilde{\vec{f}_0}(\vec{B}) + \widetilde{V}(B^2)
\ \widetilde{\vec{f_{\ }}}(\vec{B}) \>.
\eeq
Now we introduce the scalar function $\widetilde{g}$
\bmini
\widetilde{\vec{f_{\ }}}(\vec{B}) = {\vec{B} \over B^2} \ \widetilde{g}(B^2)
\eeeq
to obtain
\beeq
\vec{\nabla}_B^2 \widetilde{\vec{f_{}}}(\vec{B}) \>=\> 4\  
\vec{B} \ \widetilde{g}''(B^2) \>, \qquad 
\widetilde{g}'' \>=\> {d^2 \widetilde{g} \over d(B^2)^2} \>.
\emini
Finally, defining 
\beql{hache}
\widetilde{h}(B^2) \equiv \widetilde{g}(B^2) + 2\pi i = \vec{B} \cdot 
\widetilde{\vec{f_{\ }}}(\vec{B}) + 2\pi i\>,
\eeq
we represent (\ref{EDf}) in the form
\bminiG{EDh}
4i \kappa \ \widetilde{h}''(B^2) + {1 - \widetilde{V}(B^2) \over B^2}
\ \widetilde{h}(B^2) = 0 \>.
\eeeq 
This is a linear second order differential equation. 
The corresponding boundary conditions are
\beeq\label{bound.cond.zer}
\widetilde{h}(0) &=& 2\pi i \>,  \\ 
\label{bound.cond.inf}
\widetilde{h}(\infty) &=& 0 \>.
\emini
Indeed, (\ref{bound.cond.zer}) follows from the convergence at $B^2=0$ of the 
integral term in 
\beql{h-tilde}
\widetilde{h}(B^2)\  \equiv\ 2\pi i\ + \ \vec{B} \cdot  \int d^2 \vec{U} 
\ e^{-i\vec{B}\cdot \vec{U}}
\vec{f}(\vec{U}) \>.
\eeq
This in turn proceeds from the behaviour of $\vec{f}(\vec{U})$ 
when $\vec{U} \to \vec{0}$ or $\vec{U} \to \infty$ 
which may be inferred using (\ref{IntegralEquation}) 
and (\ref{ang.constraint1}) namely
\bmini
& \vec{f}(\vec{U})\  &\ \mathrel{\mathop \sim_{U \to 0}}\  \vec{f_0}(\vec{U}) 
\ \mathrel{\mathop \sim_{U \to 0}} \ \frac{\vec{U}}{U^2} \>; \\
& \vec{f}(\vec{U}) \ &\ \mathrel{\mathop \sim_{U \to \infty}}\  
\frac{i\vec{f_0}(\vec{U})}{\kappa U^2} \ \mathrel{\mathop \sim_{U \to \infty}}
 \ \frac{i\pi\vec{U}}{\kappa U^4} \int_{U^2}^{\infty} dQ^2 \  V(Q^2) \>.
\emini
The second boundary condition follows from the fact that
in the $B\to\infty$ limit, $\widetilde{V}(B^2)$ vanishes, and 
$\widetilde{\vec{f_{\ }}}(\vec{B})$ tends to 
$\widetilde{\vec{f_{0}}}(\vec{B})$. 
Plugged into (\ref{fzero}) and (\ref{hache}), 
this results in (\ref{bound.cond.inf}).

\paragraph{The spectrum.}

Going to ${B}$-space in the expression for the radiation density 
(\ref{U-spectrum}), we obtain 
\beq
\omega {dI \over d\omega \ dz} =\left. {2 \alpha \over\pi^2\lambda}\ 
\Re \int {d^2\vec{B} \over (2 \pi )^2}\> \widetilde{\vec{f}_0}(\vec{B}) \cdot
\widetilde{\vec{f}\ }(-\vec{B}) \ \right |_{\kappa}^{\kappa = 0} \>.
\eeq
The subtraction term ($\kappa =0$) given by the $B$-image 
of the trivial solution (\ref{kapzersol}) 
corresponds to $\widetilde{h}(B^2)=0$. 
Bearing this in mind, we substitute (\ref{fzero}) and make use 
of (\ref{hache}) to arrive at
\beql{B-spectrum}
\omega {dI \over d\omega \ dz} = {2 \alpha \over \lambda \pi} \  \Re 
\int {dB^2 \over 2 \pi i}\> {1 -
\widetilde{V}(B^2) \over B^2} \ \widetilde{h}(B^2) \>.
\eeq
Now, invoking the differential equation (\ref{EDh}) 
leads to a surprisingly simple result:
\beql{kappa-spectrum}
\omega {dI \over d \omega \
dz} = {4 \alpha \over \lambda \pi^2} \ \Re \left \{ \kappa \
\widetilde{h}'(0) \right \} \>.
\eeq
Thus the determination of the spectrum is equivalent to the following 
mathematical problem: calculate $\widetilde{h}'(0)$ for $\widetilde{h}$ 
the solution of (\ref{EDh}). 

\subsection{Solution in the small $\kappa$ limit}
\label{smallkappa}
The previous discussion is valid for general $V$'s and therefore the 
dif\-fe\-ren\-tial equation (\ref{EDh}) 
is easily treated in the limit $\kappa \ll 1$ by applying the WKB method. 
The (appropriately normalized) WKB-solution reads
\beql{WKB}
\widetilde{h}(B^2) \>\simeq\> 2\pi i \left[\,\frac{-\widetilde{V}'(B^2)\, B^2}
{1-\widetilde{V}(B^2)}\,\right]^\quart 
\exp{\left \{  -\sqrt{\frac{i}{4\kappa}} \int_{0}^{B^2} dB'^2 
\sqrt{\frac{1-\widetilde{V}(B'^2)}{B'^2}}
\right \} } \>.
\eeq
As long as $(1-\widetilde{V}(B^2))/B^2\approx-\widetilde{V}'(B^2)$ 
has a finite  $B^2\to 0$ limit, 
this solution can be applied down to $B^2=0$.
This corresponds to the case when the mean squared transverse momentum 
$\lrang{q_{\bot}^2}$ is well defined:
\bmini
 \widetilde{V}(0) &=& \int d^2Q\> V(Q^2) \>\equiv\> 1\>; \\
 -4\widetilde{V}'(0) &=& -4\frac{d}{dB^2}\int d^2Q\> V(Q^2) 
\frac{(i\vec{B}\vec{Q})^2}{2!} =  \int d^2 \vec{Q}\>Q^2\> V(Q^2)
\>\equiv\> \frac{\lrang{q_\perp^2}}{\mu^2}\>< \> \infty\>.
\emini
Evaluating $\tilde{h}'(0)$ and substituting into (\ref{kappa-spectrum}) 
yields the radiation density
\beql{XX}
 \omega  {dI \over d \omega \ dz} \> =\> {\alpha
\over \lambda \pi} \sqrt{2 \kappa (-4 \widetilde{V}'(0))} = {\alpha
\over \pi} \sqrt{\frac{\lrang{q_\perp^2}}{\lambda E^2}\, \omega}  \>\>; \qquad
\kappa \ll 1\,,
\eeq
which coincides with the original Migdal's result~\citd{M1}{Ter}. 
In Appendix \ref{gaussian} we make a closer contact with the derivation 
of Migdal, based on the random walk picture. 

In the case of interest here of Coulomb interactions,  
the WKB approximation (\ref{WKB}) is not suitable, since $\widetilde{V}'(B^2)$
developes the logarithmic singularity near $B^2=0$. 
This case is dealt with in Appendix B, the result of which 
is corroborated by a more rigorous calculation of $\widetilde{h}'(0)$ 
due to Chadan, Martin and Stubbe \cite{CMS}.
The final answer reads
\beql{spectrum(kappa)}
\omega  {dI \over d \omega \ dz} \>=\> {\alpha
\over \lambda \pi} \sqrt{\kappa \ \ln {1 \over \kappa}} \>\>; \qquad
\kappa \ll 1\,.
\eeq

\mysection{Conclusion}
\label{conc}
The aim of this paper has been to revisit the LPM effect in QED. 
The suppression of the radiation spectrum due to destructive interferences 
between radiation amplitudes induced by multiple scattering on static Coulomb 
centres has been studied. In the soft photon approximation and in the limit of 
large electron energy and infinite medium, the radiation spectrum has been 
shown to depend on the single parameter 
$\displaystyle{\kappa = {\lambda \mu^2 \over 2} {\omega \over E^2}}$ 
which characterizes the coherent nature of the effect: the LPM suppression 
appears when $\kappa < 1$.

The present approach corrects the derivation given in \cite{BDPS} for QED. 
The result for the spectrum is only slightly different from \cite{BDPS} within 
a minor change in the logarithmic factor. 

The dominant interference terms are due to centres separated by a distance 
of the order of the coherence length 
$\displaystyle{\nu \lambda \equiv \frac{1}{\sqrt \kappa} \lambda}$ 
much larger than the mean free path $\lambda$, which corresponds to large 
formation times of the radiated photon. 
The spectrum is determined by photon angles which become as large as 
$\displaystyle{u^2=U^2 \frac{\mu^2}{E^2} \sim \frac{1}{\sqrt \kappa} 
\frac{\mu^2}{E^2}}$, compared to a typical scattering angle of order 
$\displaystyle{\frac{\mu^2}{E^2}}$. The fact that $U^2$ becomes as large as 
$\frac{1}{\sqrt \kappa}$ means that the region $B^2 \sim \sqrt \kappa \ll 1$ 
is the dominant region in ``impact parameter'' space. 
In case the Fourier transform 
of the normalized scattering cross section $\widetilde{V}(B^2)$ has a finite 
derivative $\widetilde{V}'(0)$, corresponding to a scattering potential 
decreasing faster than $1/Q^2$ at large $Q$, the radiation spectrum 
is proportional to $\sqrt{\lrang{q_{\bot}^2}}=\sqrt{-4\widetilde{V}'(0) \mu^2}$. 
For a potential which is Coulombic at short distances 
$\displaystyle{\widetilde{V}'(B^2) \simeq -\frac{1}{4} \ln \frac{1}{B^2}}$ 
and the logarithm gives an enhancement factor 
$\displaystyle{\sqrt{\ln \frac{1}{\sqrt\kappa}}}$ in the radiation spectrum. 
For a general potential (\ref{kappa-spectrum}) gives the radiation spectrum 
in terms of the solution of a ``Schr\"{o}dinger'' equation (\ref{EDh}).

Finally, the present procedure can be naturally generalized to high temperature 
QCD in the spirit of \cite{BDPS}. The result of \cite{BDPS} for QCD should 
be modified accordingly, which gives for the radiated gluon spectrum
\beql{QCD}
 \omega  {d I^{(\mbox{\scriptsize QCD})} \over d \omega \ dz} 
\>=\> \frac{3\alpha_s}{2\pi} \frac{C_R}{\lambda_g} 
\sqrt{\kappa_{\mbox{\scriptsize QCD}} 
\ \ln {1 \over \kappa_{\mbox{\scriptsize QCD}}}} \nn \>,
\eeq
where $C_R=C_F$ ($N_c$) for a fast propagating quark (gluon). 
This result is valid for $\kappa_{\mbox{\scriptsize QCD}} 
= \lambda_g \mu^2 /\omega \ll 1$, 
where $\lambda_g$ is the mean free path of the gluon.

The derivation given in the present work may be extended to finite length 
media \cite{BDMPS}, which is important for future phenomenology. Last, since 
the method of derivation of the radiative
energy loss induced by multiple scattering is independent of the detailed 
form of the scattering
potential, provided it satisfies some requirements at large momentum transfer, 
a generalization to cold nuclear matter seems to be possible \citd{Levin}{BDMPS}.

\vspace{1.5 cm}
\noindent
{\bf\large Acknowledgement}

Discussions with A.~Martin and K.~Chadan are kindly acknowledged.

This research is supported in part by the EEC Programme "Human Capital
and Mobility", Network "Physics at High Energy Colliders", 
Contract CHRX-CT93-0357.

\appendix
\mysection{Gaussian interaction in the limit $\kappa \ll 1$}
\label{gaussian}
For comparison it is instructive to relate our derivation of the radiation 
spectrum to the one originally performed by Migdal \citd{M1}{Ter}. 
Instead of averaging over transverse momenta and longitudinal coordinates 
as described in (\ref{averaging}) a Gaussian probability density \cite{Shulga} 
is used as a solution of an underlying Fokker-Planck equation \citd{M1}{Ter} , 
which reflects the assumed random walk nature of multiple scattering. 
When fixed time steps equal to $\lambda$ are considered the normalized density 
reads
\beql{gaussian-averaging}
\prod_l \frac{d^2 \vec{q}_{l\bot}}{\pi \mu ^2} 
\exp(- \frac{\vec{q}_{l\bot}^{\ 2}}{\mu ^2}) \>,
\eeq
where $\mu ^2$ is identified as the squared average scattering transverse 
momentum, $\mu^2 \equiv \ <\vec{q}_{\bot}^2>$. 
\par
Starting from (\ref{U-SpectrumFactLPMBH}), and averaging with 
(\ref{gaussian-averaging}), one gets (\ref{exactLPM}) where $V$ is now 
the Gaussian interaction
\beq
V(Q^2)=\frac{1}{\pi} \exp(-Q^2) \>,
\eeq
and the function $\psi (U^2)$ is just given by the phase 
$\psi (U^2)=\exp(i \kappa U^2) \simeq 1 + i \kappa U^2$ for $\kappa \ll 1$ 
which indeed coincides with (\ref{psi}). In order to obtain the soft radiation 
spectrum we follow the steps of section \ref{smallkappa} noting that in 
${B}$-space $\widetilde{V}'(0)=-1/4$. The result is already stated 
in (\ref{XX}). The absence of a logarithmic dependence on $\kappa$ 
is qualitatively understood since scattering at large transferred momentum 
is exponentially suppressed . \par
The averaging prescription of a random walk (\ref{gaussian-averaging}) 
is also used by Blankenbecler and Drell \cite{BD}, but expressed in terms of 
the transverse electric field off which the charged particle scatters.

\mysection{The spectrum for the Coulomb case in the limit $\kappa \ll 1$}
\label{exactcalc.}
Here we derive the value of $\widetilde{h}'(0)$ in the Coulomb case.
The Fourier transform of $V(Q^2)$ is
\beal{potentialV}
\widetilde{V}(B^2) = \int  {d^2\vec{Q} \over \pi} \ {e^{-i \vec{B} \cdot
\vec{Q}} \over (Q^2 + 1)^2} = B \ K_1(B) \nn \\
\mathrel{\mathop \simeq_{B^2 \ll 1}} 1 - {B^2 \over 4} \ln {1 \over B^2} +
O(B^2) \>,
\eea
with the modified Bessel function $K_1$ \cite{AS}.\\
The differential equation (\ref{EDh}) becomes
\beql{EDCoul}
4i\kappa \ \widetilde{h}''(B^2) + {1 - B K_1(B) \over B^2}\ 
\widetilde{h}(B^2)
 = 0 \>.
\eeq
For $B^2 \ll 1$, $\widetilde{h}$ satisfies the approximate equation
\beql{EDAppr}
\widetilde{h}''(B^2) \mathrel{\mathop \simeq_{B^2 \ll 1}} -\ {1 \over      
16i\kappa}\ \ln \left ( {1 \over B^2} \right )\  \widetilde{h} (B^2) \>.
\eeq
The approximate solution
\beql{Appr.Sol.1}
\widetilde{h}(B^2)
\simeq C \exp{\left [ \pm \sqrt{{i \ln (1/B^2) \over 16
\kappa}} B^2 \right ] } 
\eeq
may  be seen, by implementation in (\ref{EDAppr}), to be valid in the
restricted region
\beql{Validity1}
\sqrt {{\kappa \over \l ^3}} \ll B^2 \ll 1 \>, \qquad 
 \l \equiv \ln {1 \over \sqrt{\kappa}} \gg  1  \>.
\eeq
The constant $C$ and the sign of the exponent in (\ref{Appr.Sol.1}) are fixed
by continuity when looking at the asymptotic forms of the solution 
$\widetilde{h}$ of (\ref{EDCoul}). For $B^2 \gg 1$ \cite{AS}
\beql{Appr.Sol.2}
\widetilde{h}(B^2) \mathrel{\mathop  \sim_{B^2 \gg 1}} C' B \ H_1^{(1)}
\left [ {1 + i \over 4 \sqrt{2}} \sqrt{{B^2 \over \kappa}} \right ]  
\mathrel{\mathop \sim_{B^2 \gg 1}} C'' \sqrt{B} \ \exp \left [ {-1 + i \over 4
\sqrt{2}} \sqrt{{B^2 \over \kappa}} \right ] \>,
\eeq
which satisfies $\widetilde{h}(\infty ) = 0$. The constants $C'$
and $C''$ are of order unity. \par

Extending the forms (\ref{Appr.Sol.1}) and (\ref{Appr.Sol.2}) to their limit 
of validity $B^2
\sim 1$, it is clear that when $\kappa \ll 1$, we have to choose the minus sign
in (\ref{Appr.Sol.1}). \par

For $B^2 \to 0$, the solution of (\ref{EDAppr}) may be written as
\beql{As.Form-2}
\widetilde{h}(B^2) \mathrel{\mathop \simeq_{B^2 \to 0}} 2i \pi +
\widetilde{h}'(0) B^2 - {\pi \over 16} \ {B^4 \over \kappa} \ln \left (
{1 \over B^2} \right ) +O(B^4)  \>.
\eeq
This expansion is valid for
\beq
B^2 \ll \left| \frac{\kappa \,\widetilde{h}'(0)}
{\ln \left(\kappa |\tilde{h}'(0)|\right)} \right| \>. 
\eeq
Assuming that $\kappa |\widetilde{h}'(0)| \hbox{\lower.7ex\hbox{$\sim\atop 
\kappa \ll 1$}} \sqrt{\kappa \l}$ (this
will be checked a posteriori in (\ref{0-derivative1})), (\ref{As.Form-2}) 
is valid for
\beql{Validity2}
B^2 \ll \sqrt{{\kappa \over \l}} \>.
\eeq
Using (\ref{Validity1}) and (\ref{Validity2}), it appears that the forms 
(\ref{Appr.Sol.1})
and (\ref{As.Form-2}) must coincide in the region
\beql{B-region}
\sqrt{{\kappa \over \l ^3}} \ll B^2 \ll \sqrt{{\kappa \over \l }} \>.
\eeq
Thus we have
\beal{0-derivative1}
C &=& 2i \pi \>,\nn \\
\widetilde{h'}(0) &=& (1 - i) {\pi \over 4} \sqrt{{2\l \over \kappa}}\>.
\eea
The final result for the radiation spectrum is obtained from 
(\ref{kappa-spectrum})
\beq
\left . \omega  {dI \over d \omega \ dz} \right |_{\kappa \ll 1} = {\alpha
\over \lambda \pi} \sqrt{\kappa \ \ln {1 \over \kappa}} \>.
\eeq

The dominant impact parameter region which determines this spectrum is given by
\beql{Bcontribution}
B^2 \sim \sqrt \kappa \>,
\eeq
corresponding to
\beql{Ucontribution}
 U^2 \sim  {1 \over \sqrt \kappa} \ \gg \ 1 \>.
\eeq

\newpage
\vbox to 2 truecm {}

\def\labelenumi{[\arabic{enumi}]}
\noindent
{\bf\large References}
\ben

\item\label{BDPS} R.~Baier, Yu.L.~Dokshitzer, S.~Peign\'e and D.~Schiff, 
\pl{345}{277}{95}.

\item\label{LP}  
  L.D.~Landau and I.Ya.~Pomeranchuk,
  {\em Dokl.~Akad.~Nauk SSSR}\/  \underline {92} (1953) 535, 735.

\item\label{M1}  
  A.B.~Migdal, \pr{103}{1811}{56}; and references therein.

\item\label{Ter}
  M.L.~Ter-Mikaelian, {\it High Energy Electromagnetic Processes 
in Condensed Media}, John Wiley \& Sons, NY, 1972.

\item\label{SLAC}
P.L.~Anthony et al., \prl{75}{1949}{95}

\item\label{BD} R.~Blankenbecler and S.D.~Drell, preprint SLAC-PUB-6944 T/E 
(September 1995).

\item\label{GW} M.~Gyulassy and X.-N.~Wang, \np{420}{583}{94};
 M.~Gyulassy, X.-N.~Wang and M.~Pl\"{u}mer, \prD{51}{3236}{95}.

\item\label{LL} L.D.~Landau and E.M.~Lifshitz, {\it Quantum Electrodynamics}, 
Course of Theoretical Physics, Vol.~4, Pergamon Press. 

\item\label{Levin} E.M.~Levin, preprint CBPF-NF-061/95.

\item\label{LR}    E.M.~Levin and M.G.~Ryskin, \sjnp{33}{901}{81}.

\item\label{GGS}
 V.M.~Galitsky and I.I.~Gurevich, 
 {\em Il Nuovo Cimento}\/ XXXII (1964) 396; \\
A.H.~S{\o}rensen, \zp{53}{595}{92}. 

\item\label{CMS} K.~Chadan, A.~Martin, and J.~Stubbe, preprint CERN-TH.96/01.

\item\label{BDMPS} 
 R.~Baier, Yu.L.~Dokshitzer, A.H.~Mueller, S.~Peign\'e and D.~Schiff, 
in preparation. 
 
\item\label{Shulga}
N.V.~Laskin, A.S.~Maz\-ma\-nish\-vi\-li and N.F.~Shul'ga, \spd{29}{638}{84};
N.V.~Las\-kin, A.S.~Maz\-ma\-nish\-vi\-li, N.N.~Na\-so\-nov and N.F.~Shul'ga, 
\spj{62}{438}{85};
A.I.~Akhiezer and N.F.~Shul'ga, \spu{30}{197}{87}, and further references 
therein.

\item\label{AS}
M.~Abramowitz and I.A.~Stegun, {\it Handbook of Mathematical Functions}, 
Dover Publ., New York, 1965.

\een

\end{document}